\documentclass[fleqn,usenatbib, 13pt]{mnras}

\usepackage{newtxtext,newtxmath}
\usepackage{verbatim}
\usepackage[T1]{fontenc}

\DeclareRobustCommand{\VAN}[3]{#2}
\let\VANthebibliography\thebibliography
\def\thebibliography{\DeclareRobustCommand{\VAN}[3]{##3}\VANthebibliography}

\usepackage{graphicx}	% Including figure files
\usepackage{amsmath}	% Advanced maths commands
\title[X-ray from outflow-cloud interaction]{X-ray from Outflow-Cloud Interaction and Its Application in Tidal Disruption Events}

\author[Chen \& Wang]{
Jiashi Chen,$^{1,2}$
Wei Wang$^{1,2}$\thanks{E-mail: wangwei2017@whu.edu.cn}
\\
$^{1}$Department of Astronomy, School of Physics and Technology, Wuhan University, Wuhan 430072, China \\
$^{2}$WHU-NAOC Joint Center for Astronomy, Wuhan University, Wuhan, 430072, China
}

\date{Accepted XXX. Received YYY; in original form ZZZ}

\pubyear{2022}

\begin{document}
\label{firstpage}
\pagerange{\pageref{firstpage}--\pageref{lastpage}}
\maketitle

\begin{abstract}
Tidal disruption events (TDEs) may occur in supermassive black holes (SMBHs) surrounded by clouds. TDEs can generate ultrafast and large opening-angle outflow with a velocity of $\sim$ 0.01--0.2 c, which will collide with clouds with time lags depending on outflow velocity and cloud distances. Since the fraction of the outflow energy transferred into cloud's radiation is anti-correlated with the cloud density, high density clouds was thought to be inefficient in generating radiation. In this work, we studied the radiation from the outflow-cloud interactions for high density clouds, and found that thermal conduction plays crucial roles in increasing the cloud's radiation. Up to 10\% of the bow shock energy can be transferred into clouds and gives rise to X-ray emission with equivalent temperature of $10^{5-6}$ Kelvins due to the cooling catastrophe. The inverse Compton scattering of TDE UV/optical photons by relativistic electrons at bow shock generates power-law X-ray spectra with photon indices $\Gamma\sim 2-3$. This mechanism may account for some TDE candidates with delayed X-ray emission, and can be examined by delayed radio and gamma-ray emissions.
\end{abstract}


%%%%%%%%%%%%%%%%%%%%%%%%%%%%%%%%%%%%%%%%%%%%%%%%%%

%%%%%%%%%%%%%%%%% BODY OF PAPER %%%%%%%%%%%%%%%%%%

\section{Introduction}

Tidal disruption event (TDE) was first coming out as a theoretical concept at late 1970s \citep{1975Natur.254..295H}. When a star plunges into the tidal radius of $R_t \approx R_{\rm star}(M_{\rm BH}/M_{\rm star})^{1/3}$, it will be torn apart by the strong tidal force.
After disruption, part of the star is bounded and falls back to the SMBH, releasing its gravitational potential to generate a luminous outburst in UV or optical band, which will decline with a timescale of months to years \citep{2015JHEAp...7..148K}.
Till now, over 100 TDEs have been discovered. Initially, it's expected that the emission is produced in the accretion disk, which should peak in UV and soft X-rays and contribute to a large fraction of TDE luminosity \citep{1988Natur.333..523R,1999ApJ...514..180U}.
However, most of observed TDEs are X-ray weak, and 11 of them are detected in soft X-rays \citep{2021ARA&A..59...21G}.
Some X-ray detected TDEs show delayed X-ray flare months or years after discovery. Their X-ray spectra are quite soft and can be well described by a blackbody with a temperature of $10^1-10^2$ eV \citep[e.g.,][]{2021SSRv..217...18S,2016MNRAS.463.3813H,2020NatCo..11.5876S,2021NatAs...5..510S}.
Besides, radio emission is also observed in some TDEs, which may originate from jet \citep{2011Sci...333..203B,2011Natur.476..421B} or outflow -- circumnuclear medium interaction \citep[e.g.,][]{2020SSRv..216...81A} or outflow--cloud interaction \citep{2022mnras.510.3650M}.

It is known that a TDE can generate outflows. Due to the general relativistic precession, the earlier debris will collide with subsequent infalling debris violently after passing the pericenter, which can generate a strong outflow with a kinetic energy of up to 10$^{51-52}$erg \citep{2020MNRAS.492..686L} and a velocity of 0.01--0.2 $c$ \citep{2016MNRAS.458.4250S}. Besides, after the circularization process of debris, an accretion disk with a high mass accretion rate will generate a strong outflow with a kinetic luminosity around 10$^{44-45}$ erg $s^{-1}$ \citep{2019MNRAS.483..565C}.

There may be some clouds around a SMBH. For the AGN, there exists a broad line region (BLR) made up of clouds, while a quiescent SMBH may also have cloud-like objects \citep{2020Natur.577..337C}.
In order to distinguish it from BLR as a concept in AGN, we call the presumed clouds around the quiescent SMBH or AGN with distances of $\sim 10^{-2}$ pc as ``dark clouds'' \citep{2022MNRAS.514.4406W}.
When a TDE occurs in a SMBH with surrounding clouds, the large-opening angle outflow will inevitably collide with clouds, triggering transient afterglows after TDE's outburst.  Depending on the distance between the SMBH and clouds, typically it takes months or years for TDE outflow arriving at BLR clouds or torus region, respectively. Outflow-cloud interaction drives two kinds of shocks: bow shocks at the windward side of the clouds, and cloud shocks inside the clouds \citep{1975ApJ...195..715M}.
The bow shock ahead of a cloud dissipates the outflow's kinetic energy into shock energy.
According to \citet{2021ApJ...908..197M} (appendix A therein), a fraction of $\chi^{1/2}$ of the bow shock energy will be transferred into cloud, where $\chi$ is the density ratio of cloud to outflow. For a high density ratio of $\chi \gtrsim 10^4$, only $\lesssim$1\% of the shock energy can be converted into cloud, while the rest is taken away by post-shock outflow. This is the case for clouds with densities similar to that of BLR (e.g., around $10^{10}$ cm$^{-3}$). When the cloud is impacted by the shock, the temperature of post-shock cloud sharply rises but soon drops below $\sim 10^4$ K due to cooling catastrophe. The strong ram pressure of bow shock then greatly increases the density of the post-shock cloud by several orders of magnitude (e.g., appendix F in \citealt{2021MNRAS.507.1684M}), which strongly reduces the energy transferring efficiency.
Nevertheless, this issue is ignored and the efficiency of cloud's radiation is largely overestimated in some works (e.g., \citealt{2017ApJ...843L..19M}).

In this work, we note that when an outflow interacts with a dense cloud, a large temperature gradient from post-shock outflow to the cloud will form, which triggers efficient thermal conduction. This helps transfer shock energy into the cloud, where the radiation is quite efficient. Thus, a new mechanism of energy transmission and radiation will be formed: outflow's kinetic energy -- shock energy (bow shock) -- thermal conduction -- cloud's radiation.
In addition, the collision between the TDE outflow and BLR clouds can also accelerate charged particles to relativistic ones. At the bow shock, accelerated electrons are exposed to the radiation field of the TDE, which will radiate X-rays via inverse Compton scattering (ICS).
This mechanism may explain some of the delayed X-ray emission in TDEs.
Following the outflow--torus interaction, here we explore the outflow--`dark' cloud interaction, in which the dark clouds are assumed to share similar properties to classical BLR clouds. For the first time, we considered the role of thermal conduction in radiation. In addition, we neglected the effects of magnetic fields.

\begin{table*}
	\caption{Fiducial parameters in our model.}
	\label{table1}
	\begin{tabular}{ p{2.5cm}  p{7cm} c} % four columns, alignment for each
		\hline
		Parameters & Descriptions & Fiducial values \\
		\hline
		$C_{\textrm{v}}$ & Covering factore of the cloud & 0.1 \\
		$R_{\textrm{d}}$ & Distance between AGN and cloud & $1.2\times10^{17}$ cm \\
		$\rho_{\textrm{c}}$ & Cloud density & $1\times10^{10}m_{\textrm{H}}$ cm$^{-3}$ \\
		$\rho_{\textrm{w}}$ & outflow density & - \\
		$L_{\textrm{kin}}$ & Kinetic luminosity of TDE outflow & $2.8\times10^{45}$ erg s$^{-1}$ \\
		$\dot{M}_{\textrm{w}}$ & Mass outflow rate of TDE outflow & $5M_{\sun} yr^{-1}$ \\
		$V_{\textrm{w}}$ & Velocity of TDE outflow & $3\times10^9$ cm s$^{-1}$ \\
		$L_{\textrm{TDE}}$ & Bolometric luminosity of TDE & $1\times10^{43}$ erg s$^{-1}$ \\
		$Z$ & Metallicity of cloud & \textbf{3 $Z_{\sun}$} \\
		\hline
	\end{tabular}
\end{table*}

The outline of the paper is organised as follows. In Section 2, we introduce the interaction between outflow and dark clouds and radiation processes. In section 3, we apply this mechanism to several TDE candidates with delayed X-ray flares. A brief discussion and summary are presented in Section 4.

\section{TDE outflow-cloud interactions}
\begin{figure*}
    \centering
    \includegraphics[width=\textwidth]{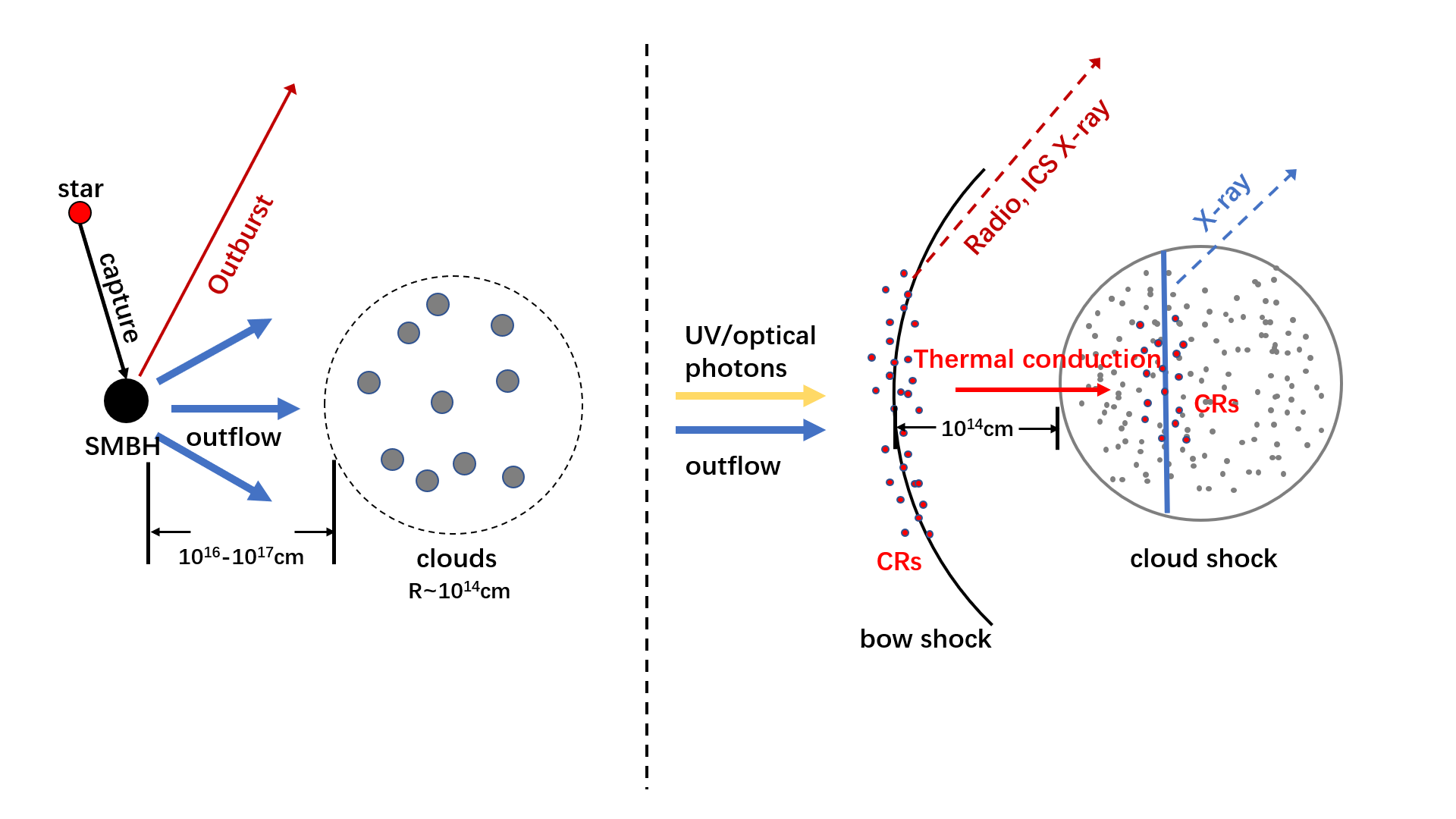}
    \caption{Sketch of the interactions between the TDE outflow and clouds. The left panel plots the overall process, a star is captured and tidally disrupted by the SMBH, inducing an outburst in optical/UV band due to a sudden increased accretion rate. The TDE outflow will reach clouds in hundreds of days. Right panel shows the outflow-cloud interaction, radiation processes and possible emissions produced.}
    \label{procedure}
\end{figure*}

A sketch of the interactions between the TDE outflow and clouds is shown in Figure \ref{procedure}. A star is captured and tidally disrupted by the gravity of SMBH,  inducing an outburst in optical/UV bands due to a suddenly increased accretion rate. An outflow is also generated during the TDE. The TDE outflow will reach and interact with clouds in hundreds of days.

Assuming ejected outflow has spherical symmetry, the outflow density is $\rho_\textrm{w}=\dot{M}_\textrm{w}/(4\pi R_d^2v_\textrm{w})\simeq1\times10^6m_{\textrm{H}}$ cm$^{-3}$, where $\dot{M}_{\rm w}$ is a mass outflow rate, $R_d$ is the distance between AGN and clouds, and $v_w$ is velocity of outflow. When the outflow encounters a dense cloud, a bow shock will form at the windward side of cloud if the outflow is supersonic. Dense cloud will be subjected to a sharp increase in pressure and a cloud shock will be driven in the cloud. The velocity of this shock is smaller than outflow \citep{1975ApJ...195..715M}. Due to the temperature difference, thermal conduction process will transfer energy from `hot' outflow to the `cold' clouds. Such a strong outflow can only last a couple of months, because both relativistic precession and accretion process are short termed.

There are no clear consensus on the mechanism determining the inner radius and effective radius of dark clouds, as well as the size of clouds. However, observations on several AGNs suggest that the cloud has a size about $10^{13\sim 14}$ cm, and at about $10^{16\sim 17}$ cm away from the SMBH, with a density around $10^8-10^{12}m_{\textrm{H}}$ cm$^{-3}$ \citep{2022MNRAS.514.1535A}. The metallicity range of clouds is $1\leq Z/Z_{\sun}\leq 5$ \citep{2013peag.book.....N}, and we take $Z=3Z_{\sun}$ in our model. Other fiducial parameters are listed in Table \ref{table1}.

\subsection{Inverse Compton cooling}

Bow shock can accelerate charged particles to relativistic ones (cosmic rays) by first-order Fermi acceleration mechanism.
%{\bf I suggest to move this part forward, at the beginning of the section 2.1:
In this work, we assume cosmic ray electrons (CRe) follow a power-law distribution:
\begin{equation}
      N_{e}(E_e) = AE_e^{-\Gamma}.
\end{equation}
%}
with a power index $\Gamma=2$. We also choose $\Gamma=2.5$ and 3.0 for comparison.

We use the precise form of the number of photons produced per unit time per unit energy from ICS \citep{1968PhRv..167.1159J}. The spectrum of Compton-scattered photons resulting from the interaction with electrons of energy $\gamma mc^2$ is:
\begin{equation}
    \frac{dN_{\gamma,ph}}{dE}=\frac{\pi r_0^2cE_{\gamma}}{2\gamma^4\beta E_{ph}^2}[F(\zeta_+)-F(\zeta_-)]
\end{equation}
where $E_{\gamma}$ is final photon energy, $E_{ph}$ is incident photon energy, $r_0$ is Bohr radius, $F(\zeta_+)$ and $F(\zeta_-)$ are equations (24)-(27) in \citet{1968PhRv..167.1159J}.

\begin{figure}
	\includegraphics[width=\columnwidth]{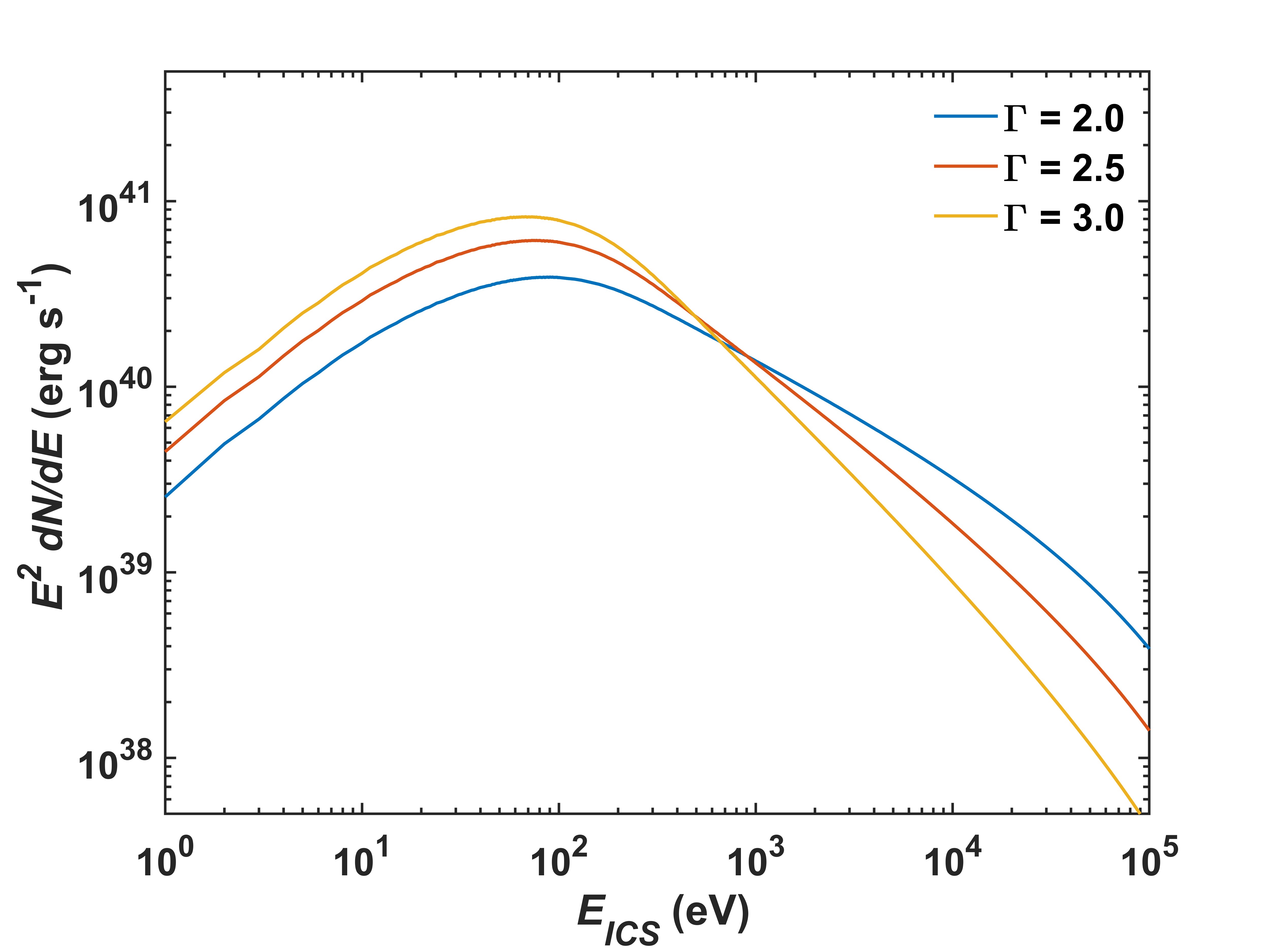}
    \caption{The spectra of inverse Compton scattering with CRes' power index $\Gamma = $ 2.0, 2.5 and 3.0. The spectrum will have a similar power-law decline after peak, and has a higher index for higher $\Gamma$. All luminosities are around $10^{41}$ erg s$^{-1}$, while it can reach $10^{43}$ erg s$^{-1}$ or even higher for stronger incident photon field or higher CRe's total energy. Seed photon field is a $2\times10^4$ K black body spectrum.}
    \label{figure1}
\end{figure}

The total Compton spectrum would be:
\begin{equation}
      \frac{\textrm{d}N_{\textrm{tot}}(E)}{\textrm{d}E}=\iint N_{e}(\gamma)\hspace{3pt}\textrm{d}\gamma(\frac{\textrm{d}N_{\gamma,ph}}{\textrm{d}E}),
      \label{equation7}
\end{equation}
 where $E$ is energy of photons after ICS, $N_e(\gamma)\textrm{d}\gamma$ is differential number of electrons, ${\textrm{d}N_{\gamma,ph}}/{\textrm{d}E}$ is the spectrum resulting from the interaction of electrons of energy $\gamma mc^2$ with an isotropic photon field.
 Figure \ref{figure1} shows the spectra of ICS with different power indexes $\Gamma$, where we use a $2\times10^4$ K black body spectrum as the seed photon field.

\subsection{Thermal conduction}

In addition to the momentum transferred from outflow to the cloud by cloud shock, energy also transferred via thermal conduction between outflow and clouds. The velocity of cloud shock is $v_{\textrm{s,c}}\simeq\chi^{-0.5}v_{\textrm{w}}$, where $\chi \equiv\rho_{\textrm{c}} /\rho_{\textrm{w}}\simeq10^4$ is the density ratio of cloud to outflow \citep{1975ApJ...195..715M}. The velocity of cloud shock is smaller than outflow, and the clouds will soon be immersed in the hot gas behind the outflow front. The windward side of cloud will ``feel'' both the thermal and the dynamical pressure of outflow, and the shock of TDE outflow would be sufficiently strong so that the initial compression at bow shock will approach 4. The temperature behind the shock front will reach $1.38\times10^5~v_{w7}^2$ K if particles are fully ionized, where $v_{w7}=v_w/(10^7$ cm s$^{-1}$) and $v_w$ is the velocity of outflow \citep{1979ApJS...41..555H}. Side and rear faces are primarily under thermal pressure, which is weaker and can be ignored.

In this paper, we try to investigate thermal conduction process and the cooling of clouds. We assume that outflow and clouds have reached a steady-state and the system is at a constant pressure equal to the ram pressure of outflow $\rho_{\textrm{w}}v_{\textrm{w}}^2$. We seek the solutions to the time-independent equations of mass, momentum and energy conservation \citep{1979ApJS...41..555H}
\begin{equation}
    \begin{aligned}
      \nabla\cdot\rho v=0\\
      \nabla\cdot(\rho \boldsymbol{vv}+\boldsymbol{\Psi}-\boldsymbol{T})=0\\
      \nabla\cdot[\textbf{v}(\frac{1}{2}\rho v^2+u)+\boldsymbol{v\cdot\Psi}+q+F]=0 ,
    \end{aligned}
    \label{equation9}
\end{equation}
where $q$ denotes the heat flux, $\nabla\cdot F=n^2\Lambda$ is the cooling term. The equations will reduce to normal hydrodynamic equations provided only that the plasma is approximately isotropic and the stress tensor may be written as $\boldsymbol{\Psi_{ij}}=p\delta_{ij}$ and $u=\textrm{Tr}\boldsymbol{\Psi}/2$ is the internal energy \citep{1979ApJS...41..555H}. The thermal conductivity in a fully ionized hydrogen plasma is \citep{1962pfig.book.....S}
\begin{equation}
    \kappa = \frac{1.84\times10^{-5}T_e^{5/2}}{\ln{\Lambda}}~\rm{ergs~s^{-1} deg^{-1} cm^{-1}},
    \label{equation10}
\end{equation}
and heat flux is $q=-\kappa\nabla T$. By solving these equations, we derive the temperature profile of the cloud. Total emission, effective temperature and density profile can be obtained from temperature profile.

\begin{figure}
	\includegraphics[width=1.1\columnwidth]{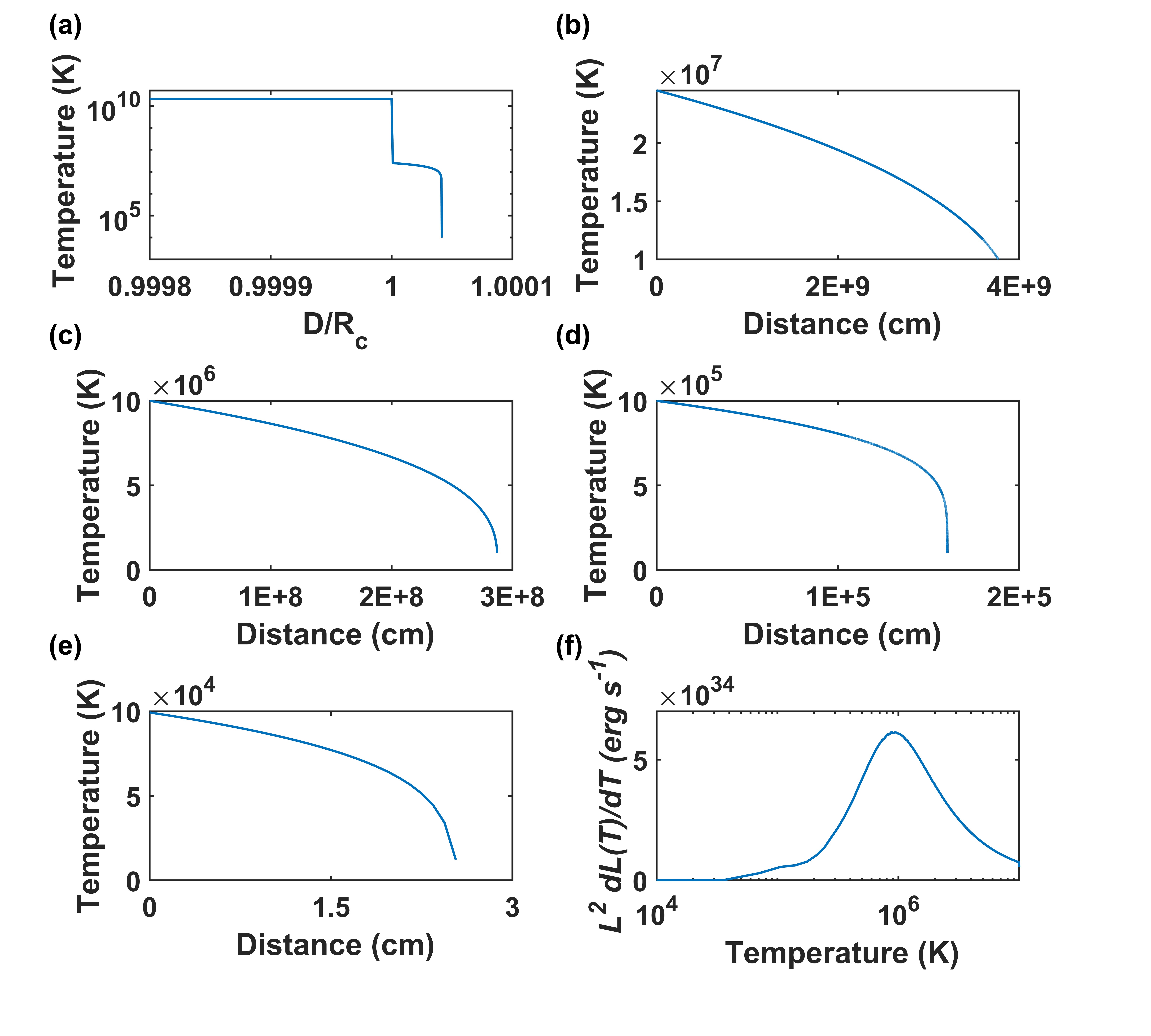}
	\includegraphics[width=\columnwidth]{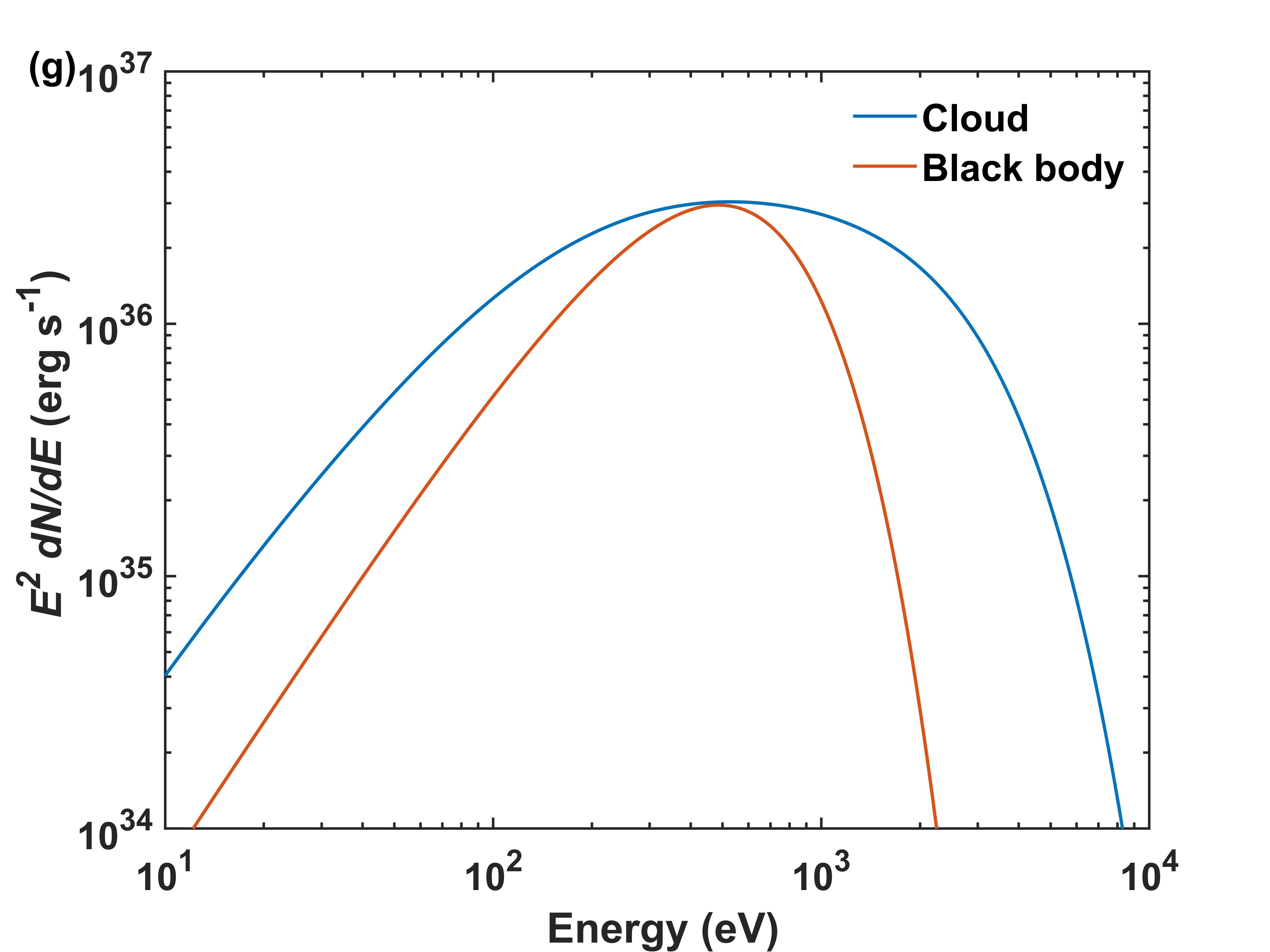}
    \caption{(a) is the overall temperature profile of cloud, where x-axis is distance in unit of cloud radius, (b), (c), (d) and (e) are the temperature profiles for cloud under our fiducial parameters, for each profile, the x-axis is the distance to the start point, where we set each star point to be 0. As a result of thermal conduction, the temperature of cloud is not uniformed, it's hot at windward side and getting cooler along the direction of outflow. Therefore, the effective temperature of cooling spectrum is not a single black body spectrum. (f) shows the luminosity at different temperatures, which peaks at about $1\times10^6$ K, indicates part of cloud around this temperature will have maximum emission. (g) gives the overall spectrum of cloud (blue), and a single $2\times10^6$ K black body spectrum (orange). The peak of cloud spectrum and the peak of black body spectrum are close, which is coincident to (f).}
    \label{figure2}
\end{figure}

The distance between the cloud and bow shock is determined by ram pressure balance condition $\rho_{\textrm{w}} V^2_{\textrm{w}} = \rho_{\textrm{refl}} V^2_{\textrm{refl}}$ \citep{1996ApJ...459L..31W}, the 'refl' means reflected and $\rho_{\textrm{refl}}=\dot{M}/(4\pi R_{\textrm{bow}}^2v_{\textrm{refl}})$, where $R_{\rm bow}$ is the distance between cloud and bow shock. Assuming that the reflected shock has a sonic-velocity and all materials hitting clouds are reflected, then $R_{\textrm{bow}}\simeq 5\times 10^{13}~\textrm{cm}$ is close to the cloud radius. Thus, we set it equals to the cloud radius for convenience. Because of high temperature $\sim 1.38\times10^5v_{s7}^2\simeq10^{10}~{\rm K}$ right behind the bow shock, saturated heat flux will reach $0.4[{2kT_e}/{(\pi m_e)}]^{1/2}n_ekT_e\simeq3\times10^{11}~{\rm erg~cm^{-2}s^{-1}}$ \citep{1977ApJ...211..135C}. The energy flux under our fiducial
parameters is $L_{\rm kin}/(4\pi R_{\rm d}^2)\simeq 1.5\times10^{10}~\rm{ergs~cm^{-2}s^{-1}}$, which is lower than the saturated heat flux and will be the maximum heat flux in our model. Thus, the heat flux is not saturated in our model, and the classical thermal conduction formula (i.e., Equation \ref{equation10}) is applicable. The thermal conductivity is $\kappa\simeq6\times10^{18}~\rm{ergs~s^{-1} deg^{-1} cm^{-1}}$ right behind the bow shock for the temperature $\sim 10^{10}$ K, and energy transfer will be efficient. The heat flux at about $10^6$ K is $q=-\kappa\nabla T\simeq 10^{9}~{\rm  {erg~cm^{-2}s^{-1}}}$, when temperature goes below $10^4$ K, the heat flux will drop rapidly to $q=10^{7}~{\rm erg~cm^{-2}s^{-1}}$, which is two orders of magnitude lower and suggests that the thermal conduction below $10^4$ K is inefficient. As shown in Figure \ref{figure2}, the temperature will decline sharply below $10^5$ K. Therefore, we ignore the thermal conduction below $10^4$ K in the following. There will be discontinuities in density and temperature at the surface of the cloud. However, across the discontinuity, the pressure is uniform and the velocity is continuous, implying no interpenetration \citep{1996AJ....111.1641T,1977ApJ...218..377W,1990ApJ...355..501B}. The discontinued area should be thin and cooling emission is negligible. Since we are interested in the spectrum of cooling process, we ignore this area.

\begin{figure}
	\includegraphics[width=1\columnwidth]{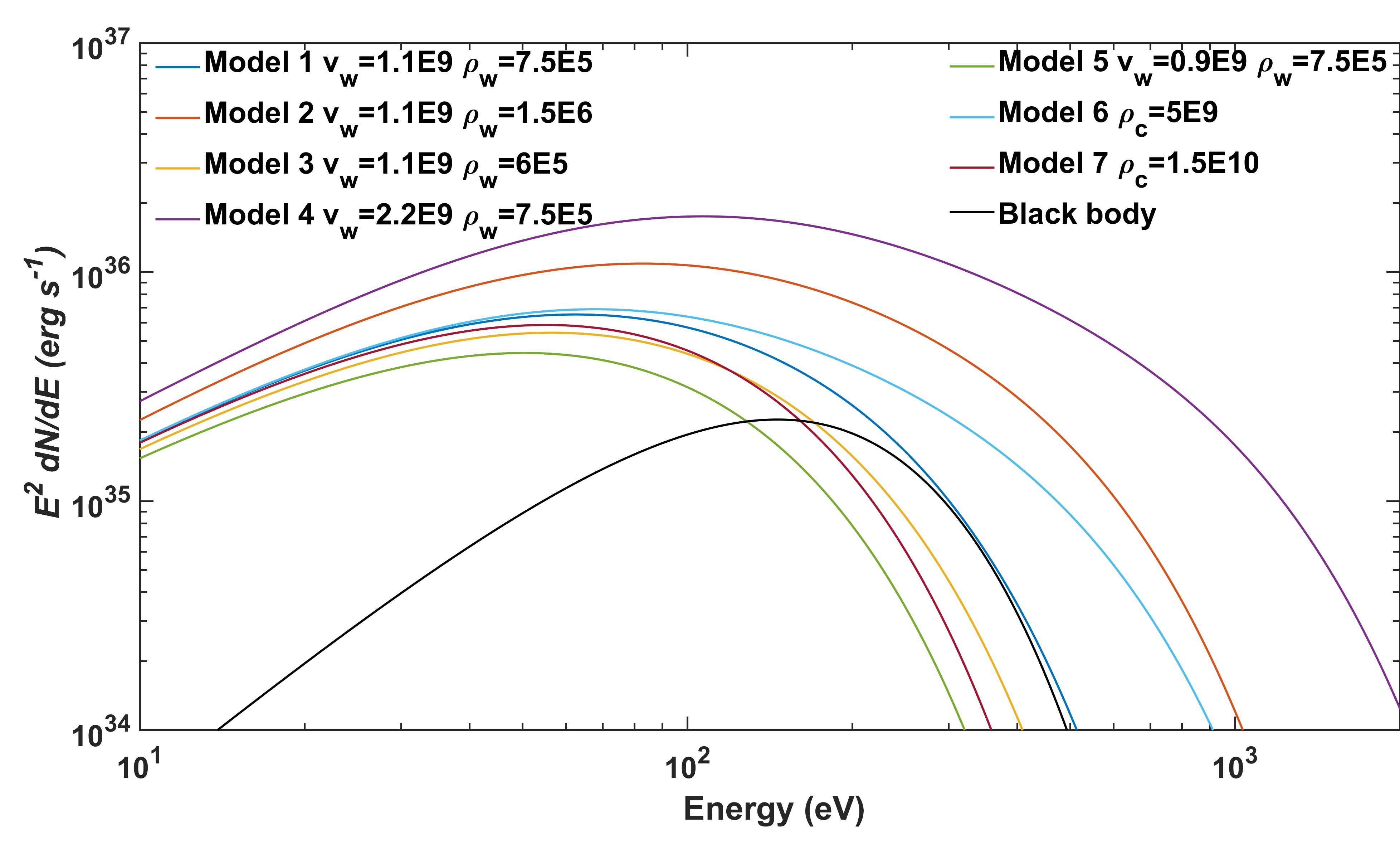}
    \caption{The effective emission spectrum for different models: model 1: $V_w=1.1\times10^9$ cm s$^{-1}$, $\rho_{\textrm{w}}=7.5\times10^5m_{\textrm{H}}$ cm$^{-3}$; model 2: $V_w=1.1\times10^9$ cm s$^{-1}$, $\rho_{\textrm{w}}=1.5\times10^6m_{\textrm{H}}$ cm$^{-3}$; model 3: $V_w=1.1\times10^9$ cm s$^{-1}$, $\rho_{\textrm{w}}=6.0\times10^5m_{\textrm{H}}$ cm$^{-3}$; model 4: $V_w=2.2\times10^9$ cm s$^{-1}$, $\rho_{\textrm{w}}=7.5\times10^5m_{\textrm{H}}$ cm$^{-3}$; model 5: $V_w=0.9\times10^9$ cm s$^{-1}$, $\rho_{\textrm{w}}=7.5\times10^5m_{\textrm{H}}$ cm$^{-3}$; model 6: same velocity and density as model 01, decrease the cloud density to $\rho_{\textrm{c}}=0.5\times10^{10}m_{\textrm{H}}$ cm$^{-3}$; model 07: same velocity and density as model 1, increase the cloud density to $\rho_{\textrm{c}}=1.5\times10^{10}m_{\textrm{H}}$ cm$^{-3}$. We use a black body of temperature $6\times10^5$ K to compare with the effective spectrum of cloud.}
    \label{figure3}
\end{figure}

Since windward shock is the main shock, we can simplify this 3-dimension question to a 1-dimension one, which is along the propagation direction of outflow. The energy equation in Equation \ref{equation9} will be
\begin{equation}
    \nabla\cdot[\frac{1}{2}\rho v^2 + \frac{3}{2}pv - \kappa\frac{\textrm{d}T}{\textrm{d}r}]+n^2\Lambda=0 .
    \label{equation11}
\end{equation}
Together with mass and  momentum equations, we can derive temperature profile behind bow shock as shown in Figure \ref{figure2}. The cooling emission for a cloud is $\int n^2(r)\Lambda(T)S\textrm{d}r$, where $\Lambda(T)$ is the cooling function \citep{1993ApJS...88..253S}. Figure \ref{figure2} shows temperature profile of cloud and emission as a function of temperature, which peaks at $T\simeq10^{6}$ K. Figure \ref{figure2} also gives the overall radiation spectrum of the cloud, which is not a single temperature black-body spectrum. The luminosity for a single cloud is about $L_{\textrm{c}}=1\times10^{37}$ erg s$^{-1}$. Equation \ref{equation11} is a function of outflow velocity and density. Thus, we change the velocity and density of outflow (the left other parameters in Table \ref{table1} unchanged) to investigate the influence of parameters on spectral properties. Figure \ref{figure3} shows the effective spectra under taking different parameters.

If thermal conduction is not considered, the efficiency of cloud shock is about $\eta_{\textrm{E}}\simeq \chi^{-1/2}=0.01$ for $\chi=10^4$, and it will drop by orders of magnitude as $\chi$ could increase by several orders of magnitude if the efficient cooling of the post-shock dense cloud gas is considered (appendix K in \citealt{2021MNRAS.507.1684M}). In contrast, the efficiency for thermal conduction is $\eta=L_{\textrm{c}}/L_{\textrm{p,c}}$, where $L_{\textrm{p,c}}=L_{\textrm{kin}}(\pi R_{c}^2)/(4\pi R_d^2)$ is the power of the outflow impacting the cloud. More detailed results of the efficiency and effective temperature under different parameters are presented in Appendix \ref{A}. The effective temperature generally is propotional to the outflow velocity and density. The efficiency will increase (up to $\sim 15\%$) as the outflow density increases, but will decrease firstly and then increase as outflow velocity increases for a given density. If the ram pressure $\rho_{\rm w}v_{\rm w}^2$ doesn't change when varying outflow velocity and density, we will get the same effective temperature.

Thus, we conclude that by means of thermal conduction, the outflow impacting on the cloud can convert $\eta \sim (5-10)\%$ of its energy into cloud radiation, which is much higher than that without thermal conduction. Considering a covering factor of $C_v$, the total X-ray luminosity from the cloud can reach $L_{\textrm{tot}}=\eta C_{\textrm{v}}L_{\textrm{kin}}\simeq 1\times10^{43}$ erg s$^{-1}$ $(\eta/0.1) (C_v/0.1) (L_{\rm kin}/10^{45})$, implying that this mechanism may account for the delayed X-ray radiation of some TDE candidates.

\section{TDEs with delayed X-ray brightening}

There are over 100 TDEs discovered up to now, and 11 of them are detected in soft X-rays, the ratio of X-ray component to UV/optical component shows dramatic variability with time, which can reach one or even higher at hundreds of days after optical peak \citep{2021ARA&A..59...21G}. In this section we choose four TDEs (i.e., ASASSN-15oi, AT2019azh, AT2018fyk, OGLE 16aaa) and try to explain the delayed brightening and spectrum of X-rays by using our outflow-cloud interaction model. The fitted parameters for the observed X-ray spectra of these TDEs are shown in Table \ref{table2}.

\subsection{ASASSN-15oi}

ASASSN-15oi was first reported by \citet{2016MNRAS.463.3813H}, which locates at a luminosity distance $D\simeq214$ Mpc with a redshift $z=0.0479$ \citep{2018MNRAS.480.5689H}. The central black hole has a mass $\sim10^6M_{\sun}$ \citep{2017ApJ...851L..47G}. ASASSN-15oi is associated with radio emission \citep{2021NatAs...5..491H}, which indicates the presence of a jet or outflow. Unlike other optically discovered TDEs, ASASSN-15oi faded rapidly in the optical and UV, and almost completely disappeared about 3 months after discovery \citep{2016MNRAS.463.3813H}. Its UV/optical spectra can be fitted by a black body, and its temperature increased from $2\times10^4$ K to $4\times10^4$ K in the first few months since discovery. In the early stage, the UV/optical luminosity declined following a $t^{-5/3}$ power-law, and $t^{-5/12}$ in later time \citep{2018MNRAS.480.5689H}. In contrast to the optical/UV luminosity evolution, the X-ray luminosity remained constant at early time and was about two orders of magnitude weaker than the UV/optical emission. But during $200\sim400$ days, the X-ray emission increased by more than an order of magnitude and had a luminosity comparable to UV/optical luminosity ($L_{\textrm{opt}}/L_{X}\sim1$ \citealt{2017ApJ...851L..47G}). The X-ray emission faded again at later time.

\begin{figure}
	\includegraphics[width=\columnwidth]{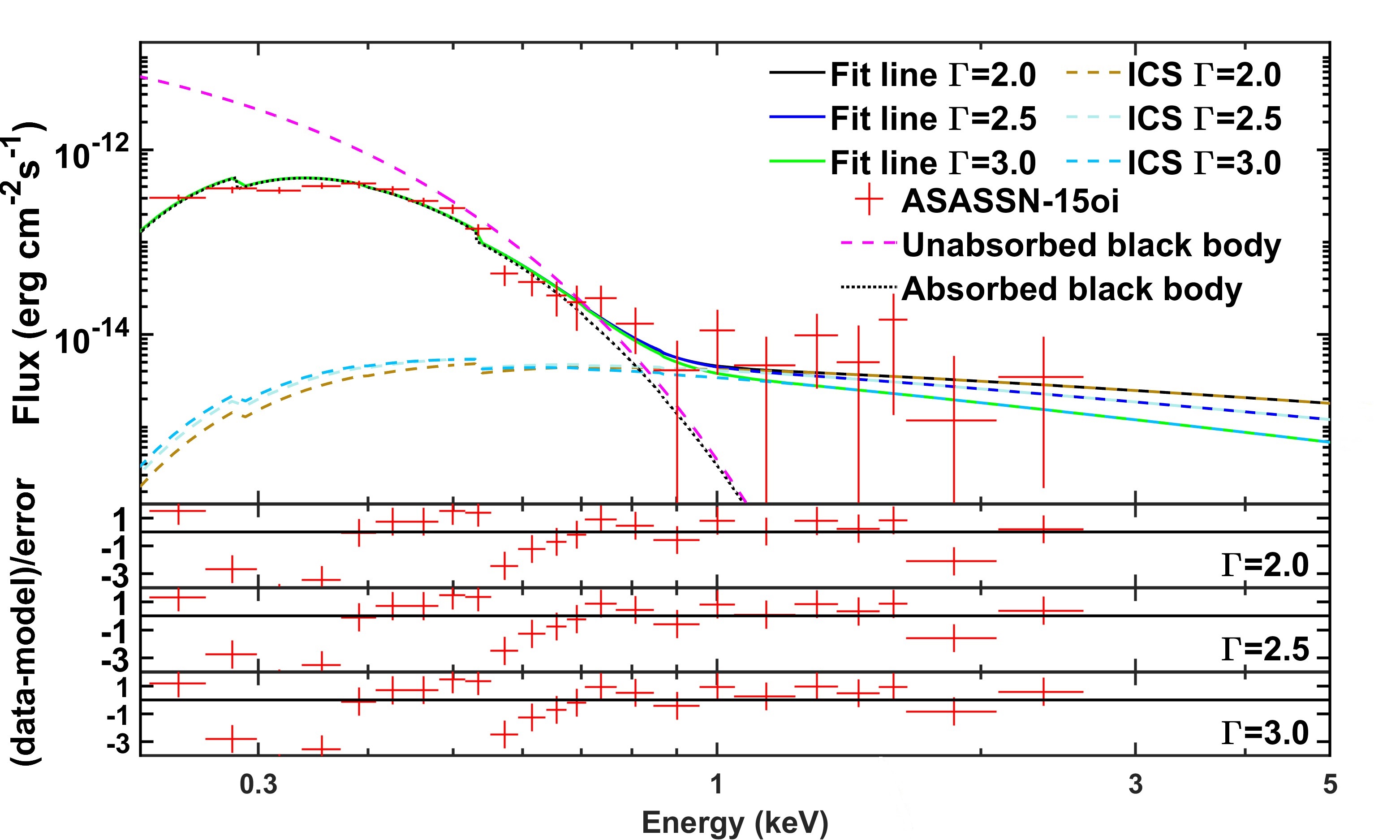}
	\caption{The overall spectrum of ASASSN-15oi. Outflow has a velocity $V_{\textrm{w}}=1.4\times10^9$ cm s$^{-1}$ and density $\rho_{\textrm{w}}=1.9\times10^7m_{\textrm{H}}$ cm$^{-3}$. We choose CRe power-law index $\Gamma =2.0,2.5$ and $3.0$.}
    \label{figure4}
\end{figure}

We focus on the observation XMM 0722160701 which was taken about 230 days after the TDE discovery. This observation gave a soft X-ray spectrum that can be fitted by a black body of $0.053\pm0.02$ keV with a flux $3.2^{+0.9}_{-1.0}\times10^{-13}$ erg cm$^{-2}$ s$^{-1}$ and a power-law part with a flux $1.5^{+0.8}_{-0.8}\times10^{-14}$ erg cm$^{-2}$ s$^{-1}$ \citep{2018MNRAS.480.5689H}. To fit the blackbody part of the spectrum, we take outflow velocity $V_{\textrm{w}}=1.4\times10^9$ cm s$^{-1}$ and density $\rho_{\textrm{w}}=1.9\times10^7m_{\textrm{H}}$ cm$^{-3}$, cloud density $\rho_{\textrm{c}}=3.6\times10^{11}m_{\textrm{H}}$ cm$^{-3}$. Figure \ref{figure4} shows overall fitting spectra with different power-law indices. We use a Galactic H${\mathrm{1}}$ column density of $5.6\times10^{20}$ cm$^{-2}$ \citep{2016MNRAS.463.3813H}, and photo-electric cross section per hydrogen in \citet{1983ApJ...270..119M}. The unabsorbed X-ray flux is about $3\times10^{-13}$ erg cm$^{-2}$ s$^{-1}$, which corresponds to a luminosity $L_X\sim10^{42}$ erg s$^{-1}$ and is close to the result in \citet{2018MNRAS.480.5689H}.

\subsection{AT2019azh}

The AT2019azh was first discovered on 2019-02-22 (MJD 58529.44) by All Sky Automated Survey for SuperNovae (ASSASN) \citep{2019TNSCR.287....1B}. AT2019azh locates at a luminosity distance $D\simeq96$ Mpc, with a redshift $z=0.022$ and a Galactic absorption column density of $N_{\textrm{H}}=4.15\times10^{20}$ cm$^{-2}$ \citep{2019arXiv191206081L}. Its central SMBH has a mass $M_{\textrm{BH}}\lesssim4\times10^6$ M$_{\sun}$.  The early-time spectra of AT2019azh have the very blue continuum that is a hallmark of tidal disruption events \citep{2021MNRAS.500.1673H}. Radio emission was detected months after the detection of UV/optical outburst \citep{2022MNRAS.511.5328G}, which indicates the existence of outflow or jet. The UV/optical luminosity declines following an exponential law with time, while the X-rays show a late brightening by a factor of $\sim$ 30-100 at around 250 days after discovery \citep{2019arXiv191206081L}. The temperature of AT2019azh was around $T\simeq10^4$ K at about 250 days after discovery. The evolution of AT2019azh's luminosity ratio of X-rays over UV/optical is similar to that of ASASSN-15oi. This ratio rose from $\sim0.01$ at the early stage to $\sim1$ at around a year after discovery.

\begin{figure}
	\includegraphics[width=\columnwidth]{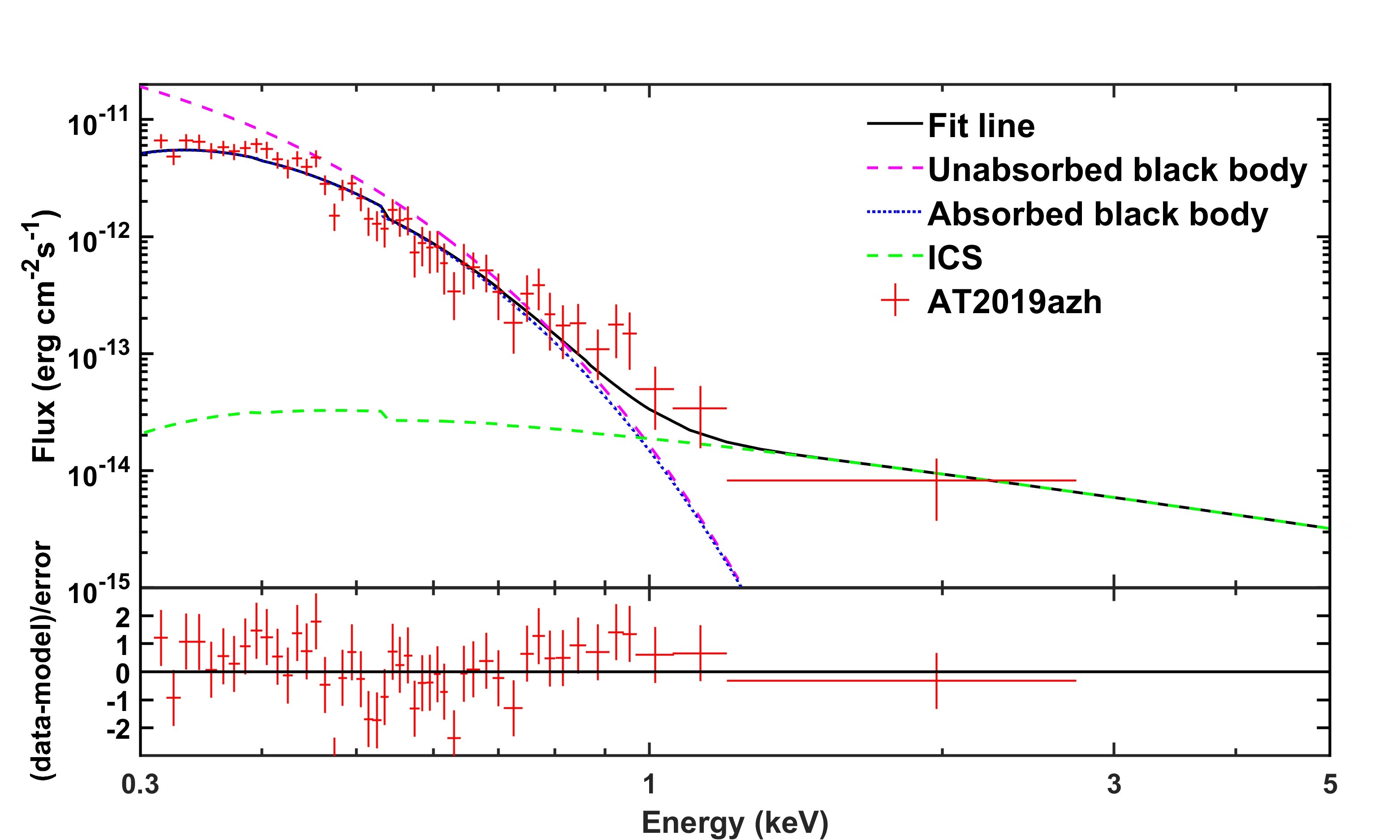}
	\caption{The fitting spectrum of AT2019azh. Outflow has a velocity $V_{\textrm{w}}=1.2\times10^9$ cm s$^{-1}$ and density $\rho_{\textrm{w}}=2.9\times10^7m_{\textrm{H}}$ cm$^{-3}$, CRe power-law index $\Gamma=3.0$.}
    \label{figure5}
\end{figure}

We study on this late-time brightening at around 250 days after the TDE discovery with the X-ray flux of $\sim5.9\times10^{-12}$ erg cm$^{-2}$ s$^{-1}$. The spectrum can be fitted by a blackbody part of $0.053^{+0.006}_{-0.004}$ keV plus a power-law part \citep{2019arXiv191206081L}. To fit the blackbody part of the spectrum, we take outflow velocity $V_{\textrm{w}}=1.2\times10^9$ cm s$^{-1}$ and density $\rho_{\textrm{w}}=2.9\times10^7m_{\textrm{H}}$ cm$^{-3}$, and cloud density $\rho_{\textrm{c}}=3.3\times10^{11}m_{\textrm{H}}$ cm$^{-3}$. Figure \ref{figure5} shows the fitting result based on the outflow-cloud interaction model.

\subsection{AT2018fyk}

The AT2018fyk was first discovered by ASASSN survey on 2018 September 8, and this transient is likely a TDE \citep{2019MNRAS.488.4816W}. AT2018fyk locates at a luminosity distance 264 Mpc, with redshift $z=0.059$ and a Galactic absorption column density $N_H=1.15\times10^{20}$ cm$^{-2}$ \citep{2021ApJ...912..151W}. The UV/optical temperature is about $4\times10^4$ K during the first several months after discovery. The UV/optical light-curve of AT2018fyk appears a smooth decline. At about 20 days after discovery, the X-ray emission increased by a factor of $\sim 10$ in 6 days and remained unchanged for months. The X-ray evolution may be explained by reprocessing process or delayed accretion \citep{2019MNRAS.488.4816W,2021ApJ...912..151W}. AT2018fyk and ASASSN-15oi have similar evolution of optical to X-ray ratio and late-time X-ray brightening and we try to fit the spectrum with our model.

\begin{figure}
	\includegraphics[width=\columnwidth]{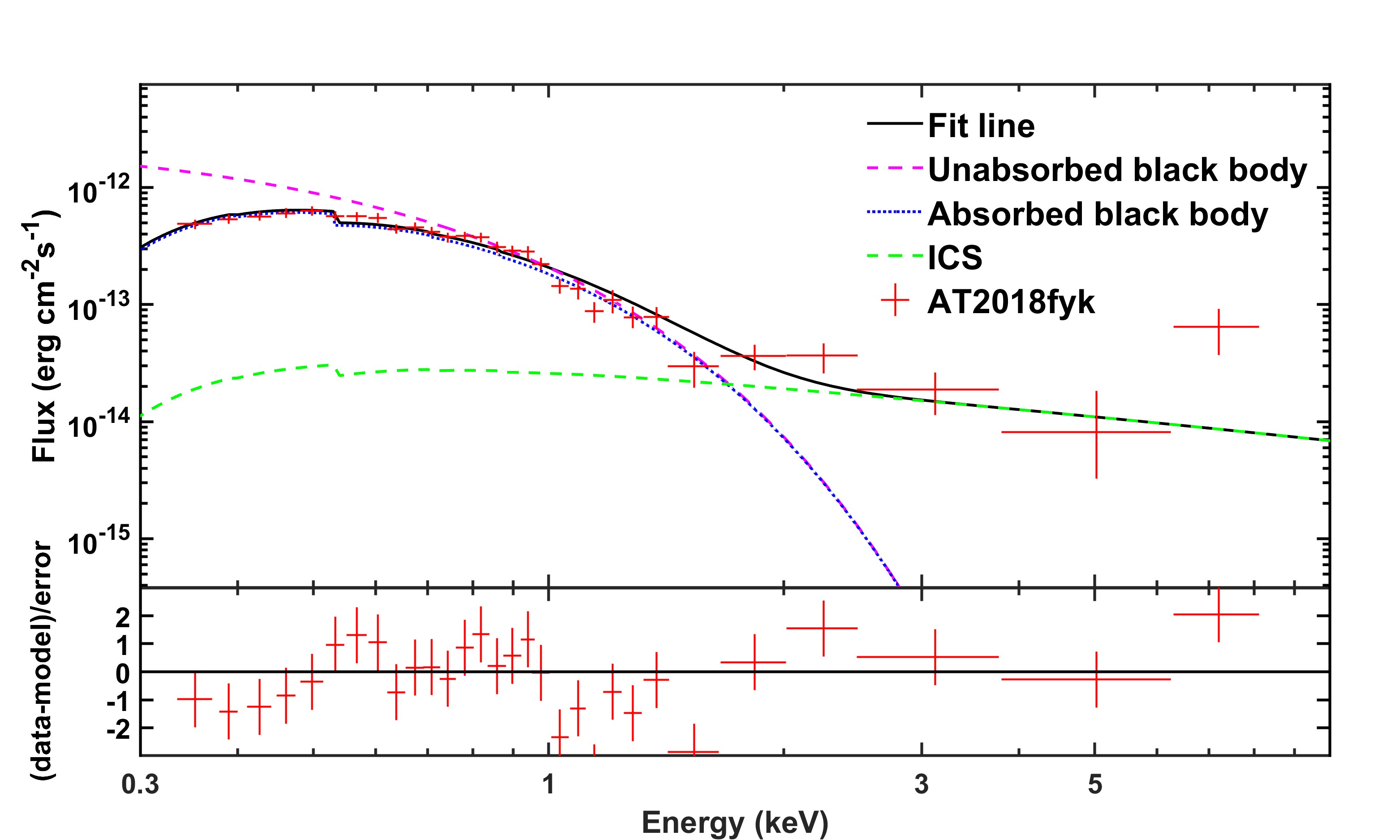}
	\caption{The fitting to the spectrum of AT2018fyk observed by XMM-Newton. In this model, outflow velocity $V_{\textrm{w}}=2.0\times10^9$ cm s$^{-1}$ and density $\rho_{\textrm{w}}=9\times10^7m_{\textrm{H}}$ cm$^{-3}$, and a power index $\Gamma=2.0$.}
    \label{figure7}
\end{figure}

AT2018fyk was observed by XMM-Newton European Photon Imaging Camera (EPIC) several times, the first observation was taken at about 92 days after discovery. The X-ray of this observation has a flux about $6\times10^{-13}$ erg cm$^{-2}$ s$^{-1}$, and can be fitted by a black body part with a temperature $kT\simeq 123$ eV and a power-law part \citep{2021ApJ...912..151W}. To fit the blackbody spectrum in x-rays, we take the outflow velocity $V_{\textrm{w}}=2.0\times10^9$ cm s$^{-1}$ and density $\rho_{\textrm{w}}=9\times10^7m_{\textrm{H}}$ cm$^{-3}$, and cloud density $\rho_{\textrm{w}}=1.4\times10^{12}m_{\textrm{H}}$. Figure \ref{figure7} shows the fitting result. We notice the spectrum has a peak at about 6-7 keV, which may be the iron K$\alpha$ emission line.

\subsection{OGLE 16aaa}

OGLE16aaa was first discovered by OGLE Transient Detection System at the center of its host galaxy on January 2, 2016, and was first reported by \citet{2017MNRAS.465L.114W}. OGLE16aaa locates at a luminosity distance $D\simeq819.4$ Mpc, with a redshift $z=0.1655$ and a Galactic absorption column density of $N_{\textrm{H}}=2.71\times10^{20}$ cm$^{-2}$. Its central SMBH has a mass $M_{\textrm{BH}}\simeq3.0\times10^6$ M$_{\sun}$ \citep{2020A&A...639A.100K}. The UV/optical light-curve of OGLE16aaa indicates a roughly power-law decay with a theoretical power index $\gamma=-5/3$. The temperature remained about $\sim2\times10^4$ K in the first hundreds of days \citep{2020NatCo..11.5876S}. There is no X-ray emission detected in early times of OGLE16aaa, and it was first detected at about 141 days after discovery by XMM-Newton. The X-ray emission rose to a peak at about 300 days after discovery and faded.

\begin{figure}
	\includegraphics[width=\columnwidth]{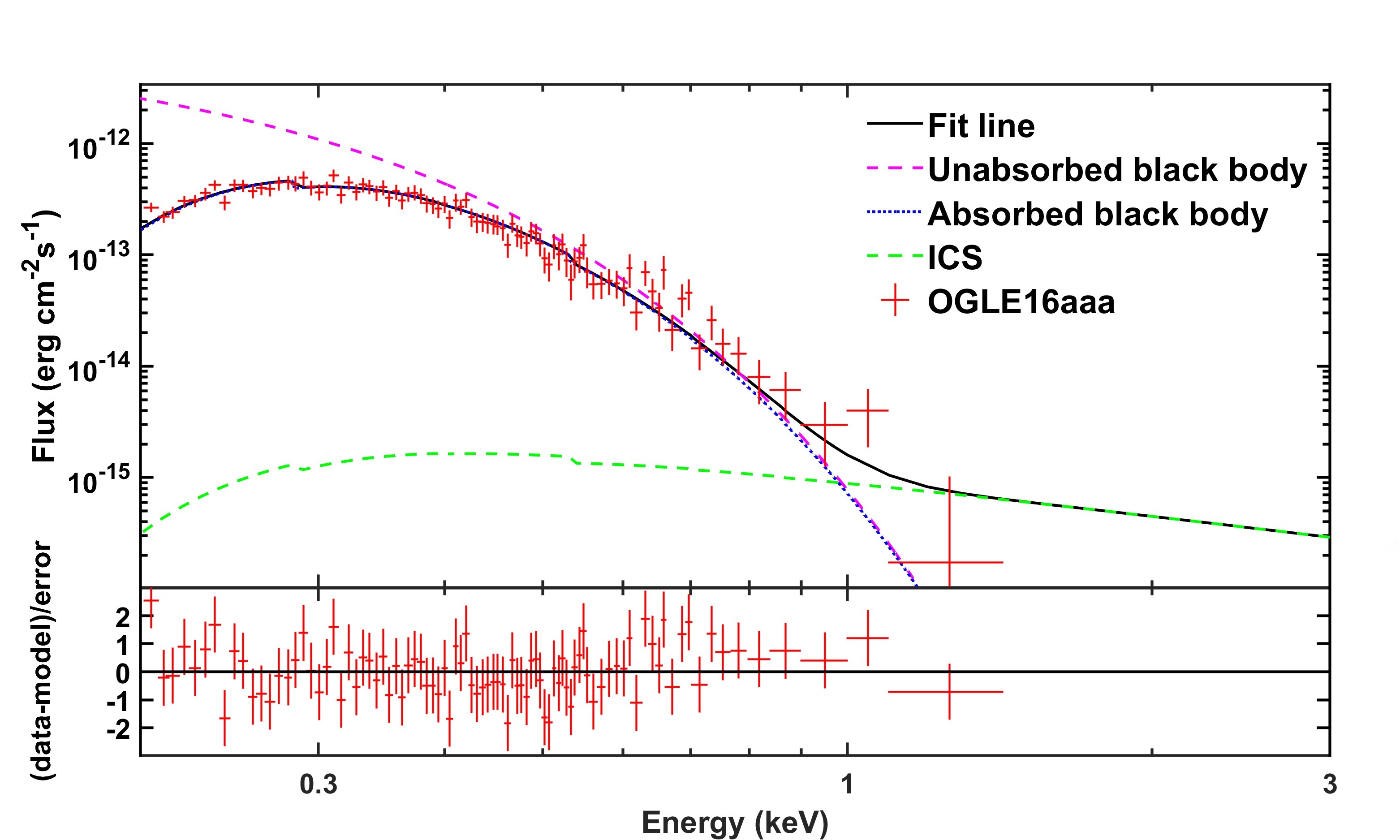}
	\caption{The fitting to the spectrum of OGLE 16aaa observed by XMM-Newton on November 30 2016 (OBSID: 0793183201). In this model, outflow velocity $V_{\textrm{w}}=1.2\times10^9$ cm s$^{-1}$ and density $\rho_{\textrm{w}}=7.5\times10^7m_{\textrm{H}}$ cm$^{-3}$, and a power-law index $\Gamma=3.0$.}
    \label{figure6}
\end{figure}

XMM-Newton observed OGLE16aaa on November 30 2016 (above 316 days after the TDE discovery) for 36.6 ks, finding a X-ray flux of $\sim 3.2\times10^{-13}$ erg cm$^{-2}$ s$^{-1}$. The spectrum can be well described by a two-blackbody model ( $\sim$ 51 eV and $\sim$ 90 eV) or a blackbody part ($\sim 60$ eV) plus a power-law part \citep{2020NatCo..11.5876S}. This late-time X-ray brightening is similar to above two TDEs. To fit the blackbody part of the spectrum, we take outflow velocity $V_{\textrm{w}}=1.2\times10^9$ cm s$^{-1}$ and density $\rho_{\textrm{w}}=7.5\times10^7m_{\textrm{H}}$ cm$^{-3}$, and cloud density $\rho_{\textrm{c}}=6\times10^{11}m_{\textrm{H}}$ cm$^{-3}$. Figure \ref{figure6} shows the fitting result. In our model, the mass outflow rate is about 38 $M_{\sun}\rm yr^{-1}$ which may be hard to achieve in the TDE based on the present simulations \citep{2019MNRAS.483..565C}.

Some other models are also proposed to explain this delayed X-ray brightening in TDEs. The late-time X-ray brightening may arise from a delayed accretion of circularized gas onto the SMBH \citep{2020A&A...639A.100K}, or possibly the X-ray is blocked by radiation-dominated ejecta in the early phase \citep{2020NatCo..11.5876S}. These models would require the enhanced accretion rate in TDEs after hundreds of days, which would dominantly contribute to X-rays not UV/optical bands. The wind-cloud interaction model would naturally produce the delayed X-ray emission properties.

%The X-rays may still come from TDE accretion, and X-ray brightening at $t\simeq 250$ days is due to a higher accretion rate \citep{2019arXiv191206081L}.
 %Our model may not suitable for OGLE16aaa.

\section{Summary and discussion}

In this paper we have developed a TDE outflow model to explain the delayed brightening of TDE X-ray emission. In our model, we argue that TDE will generate an ultra-fast outflow which can reach a velocity about 0.1$c$. The outflow will reach dark cloud about hundreds of days after launch. outflow-cloud interaction will generate two shocks: bow shock and cloud shock. Bow shock will heat the CRes to relativistic state with a power-law distribution. UV/optical photons from TDE will scatter from CRes (ICS) and generate X-ray emission. Thermal conduction between clouds and outflow will heat clouds up efficiently and the cloud will radiate like a black body. The time-delayed X-ray spectrum is composed by ICS and thermal cooling emissions.

In our thermal conduction model, we assume that there is a density discontinuity at the surface of the cloud. Theoretically, there would be a discontinuity at surface of cloud before outflow-cloud interaction, but the consequent equilibration of the pressure will smooth the density profile at the discontinuity between the outflow and ISM material \citep{1977ApJ...218..377W}. \citet{1996AJ....111.1641T} suggests a list of possible ways to disrupt the contact discontinuity, such as cloud crushing, thermal evaporation, hydrodynamical instabilities, as well as the effects caused by explosions inside outflow-driven shells and by fragmented ejecta, are evaluated. When the discontinuity is smoothed, we argue that density is still declining in a short scale and emission of this region can be ignored.

To test our model, we have analysed four TDEs that are detected with delayed X-ray brightening, relative parameters fitting the X-ray spectra are displayed in Table \ref{table2}. Because of a lack of enough observation data for these TDEs at present, we can hardly constrain the outflow parameters well yet. Anyway, the fitted values of the TDE outflow parameters would be reasonable. The numerical simulations \citep{2019MNRAS.483..565C, 2018ApJ...859L..20D} suggest that the powerful outflows from TDEs could reach a kinetic luminosity from $10^{44}-10^{46}$ erg s$^{-1}$ with the velocity of outflow in the range of $\sim 0.1- 0.2 c$, and the mass outflow rate up to $\sim 10 M_{\sun}\rm yr^{-1}$. If the velocity or density of the outflow can be obtained by multiwavelength observations \citep[e.g., ASASSN-14li with an outflow velocity of 0.2$c$, ][]{2018MNRAS.474.3593K}, we may constrain the other outflow parameters.

\begin{figure}
	\includegraphics[width=\columnwidth]{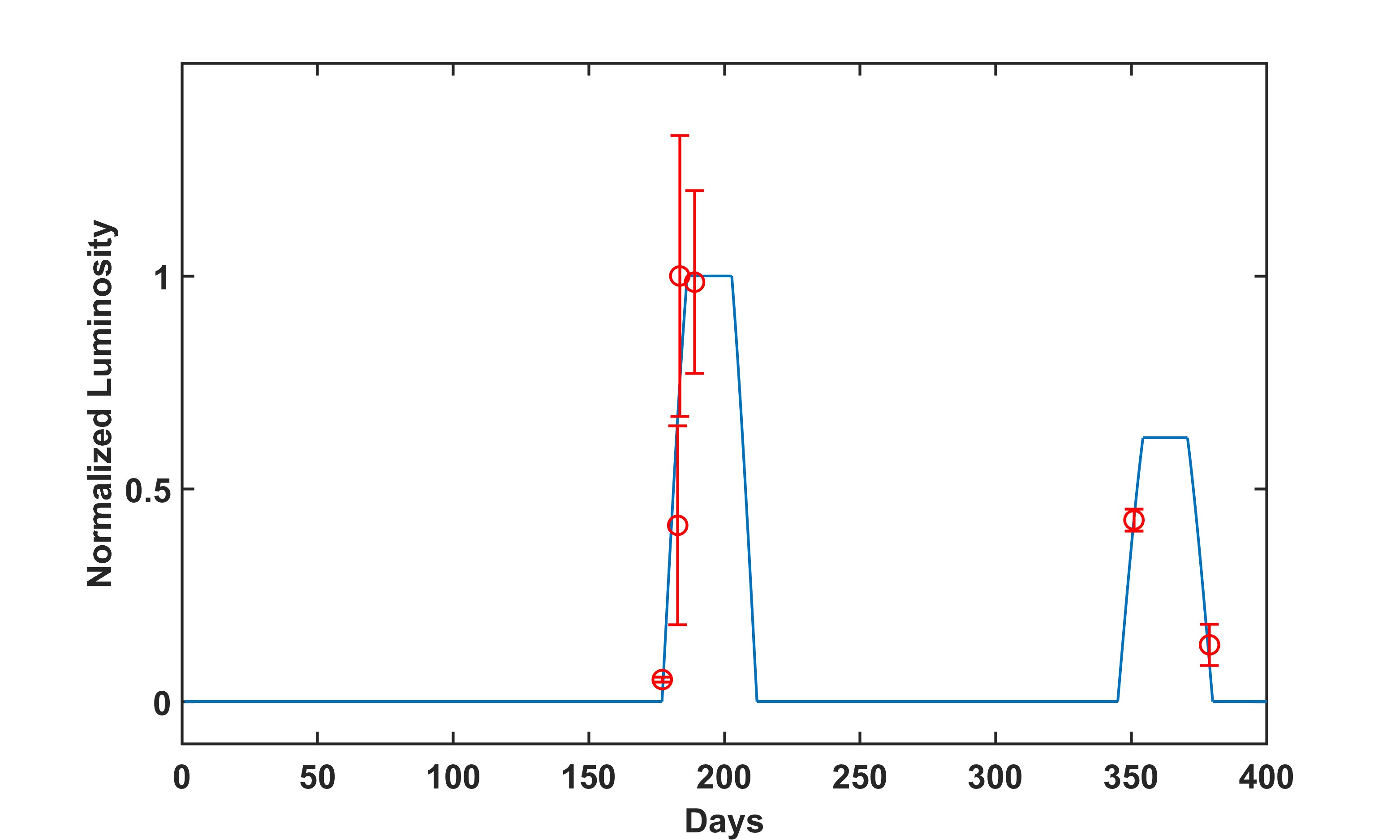}
	\caption{X-ray light curves produced by outflow-cloud interactions when there are two cloud zones near the AGN centre, where y axis is normalized luminosity in unit of the peak luminosity of outflow interacting with the first cloud zone, x axis is the time in units of days since the TDE discovery. Red markers are the normalized observed X-ray luminosity of OGLE16aaa (from \citealt{2020NatCo..11.5876S}), the blue line is the simulated luminosity evolution.}
    \label{figure8}
\end{figure}

Besides, there may exist the density (or cloud distribution) inhomogeneity along the outflow expanding routine, then the outflow-cloud interaction will also produce strong X-ray emission variations. If we assume the first cloud zone at a mean distance $R_d\simeq 5\times10^{16}$ cm from AGN centre and the second one locates at a distance $\sim 2R_d$, outflow remains its velocity after encountering with the first cloud zone and two cloud zones share similar parameters as shown in Table \ref{table1}. By using the relationship between outflow density and energy transfer efficiency in Appendix \ref{A}, we can predict the X-ray luminosity evolution generated by outflow-cloud interactions. Figure \ref{figure8} shows the X-ray emission luminosity evolution when there are two cloud zones near the AGN centre. This simulated result could explain the delayed X-ray light curves of OGLE 16aaa \citep[][Fig. 5 therein]{2020NatCo..11.5876S}, which has the variable X-ray luminosity showing a peak at around 180 days and a lower peak at about 350 days.

We demonstrate that the X-ray spectrum can be well fitted by our model, where we change outflow velocity and density to fit black body part with different effective temperatures. These delayed bright X-ray emissions, accompanied with the year-delayed radio and gamma-ray afterglows would be good candidates for future surveys \citep{2021MNRAS.507.1684M,2022MNRAS.514.4406W}. We only have limited candidates and data at present. When more TDEs and delayed emissions are discovered by X-ray telescopes in the future, fruitful information can be used to verify this scenario.

\setcounter{table}{1}
\begin{table*}
	\caption{Best fitted parameters including outflow velocities and densities, cloud densities and outflow luminosities for four TDEs}
	\label{table2}
	\begin{tabular}{ lccccc}
		\hline
		Name & Outflow velocity & Outflow density & Cloud density & Mass outflow rate & Kinetic luminosity\\
		 &  (cm s$^{-1}$)&  ($m_{\textrm{H}}$ cm$^{-3}$) & ($m_{\textrm{H}}$ cm$^{-3}$) & ($M_{\sun}\rm yr^{-1}$) & (erg s$^{-1}$) \\
		\hline
        ASASSN-15oi & $1.4\times10^9$  & $1.9\times10^7$ & $3.6\times10^{11}$  & 6.0 & $3.5\times10^{44}$ \\
        AT2019azh   & $1.2\times10^9$  & $2.9\times10^7$ & $3.3\times10^{11}$  & 9.1 & $4.1\times10^{44}$ \\
        AT2019fyk   & $2.0\times10^9$  & $9.0\times10^7$ & $1.4\times10^{12}$  & 15.3 & $1.9\times10^{45}$ \\
        OGLE16aaa   & $1.2\times10^9$  & $7.5\times10^7$ & $6.0\times10^{11}$  & 38.0 & $1.6\times10^{45}$ \\
		\hline
	\end{tabular}
\end{table*}

\section*{Acknowledgements}
We are grateful to the referee for the useful comments to improve the manuscript. The authors also thank Guobin Mou for suggesting the important role of thermal conduction in outflow-cloud interaction.
This work is supported by the National Key Research and Development Program of China (Grants No. 2021YFA0718500, 2021YFA0718503), the NSFC (12133007, U1838103).

\section*{Data Availability}
The data used in this paper were collected from the previous literatures. These X-ray spectra are public for all researchers. The calculation codes and simulated data are available by contacting with the authors.



\bibliographystyle{mnras}
\bibliography{DRAFT} % if your bibtex file is called example.bib

%\begin{thebibliography}{99}
%\bibitem[\protect\citepauthoryear{Author}{2012}]{Author2012}
%\bibitem[\protect\citepauthoryear{Others}{2013}]{Others2013}
%\end{thebibliography}

%%%%%%%%%%%%%%%%%%%%%%%%%%%%%%%%%%%%%%%%%%%%%%%%%%

%%%%%%%%%%%%%%%%% APPENDICES %%%%%%%%%%%%%%%%%%%%%
\appendix

\section{Parameter dependence of efficiency and effective temperature}\label{A}

\begin{figure}
	\includegraphics[width=\columnwidth]{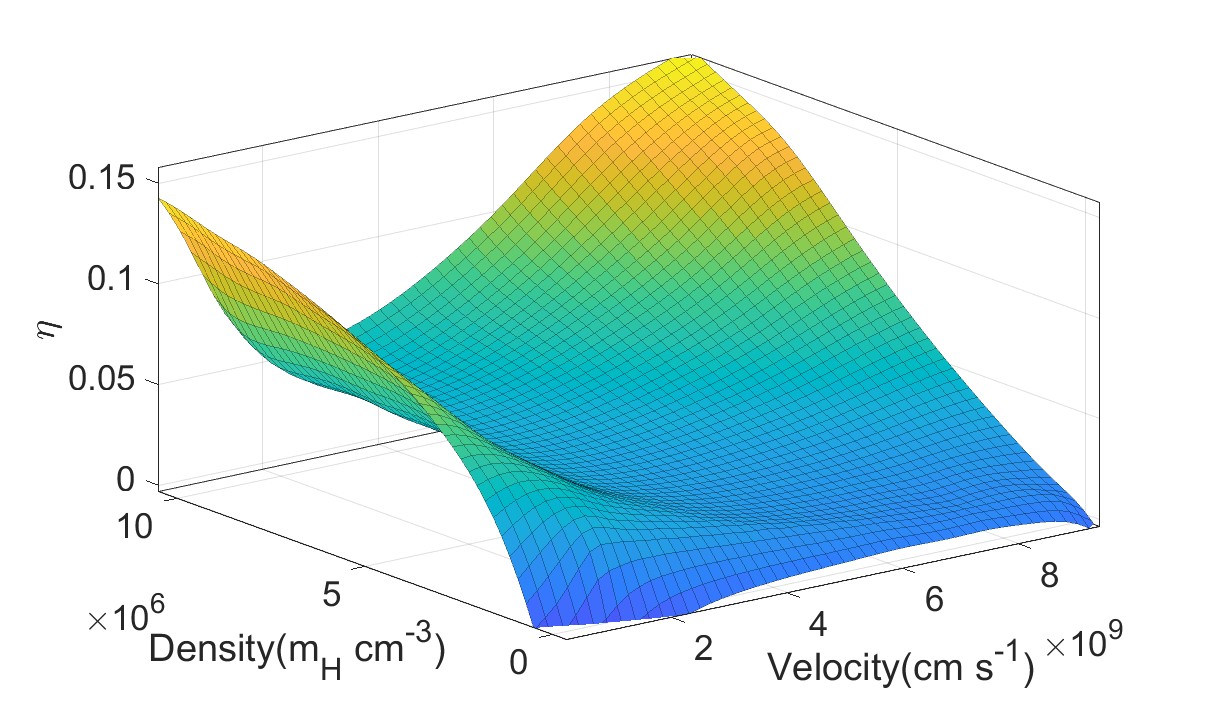}
	\includegraphics[width=\columnwidth]{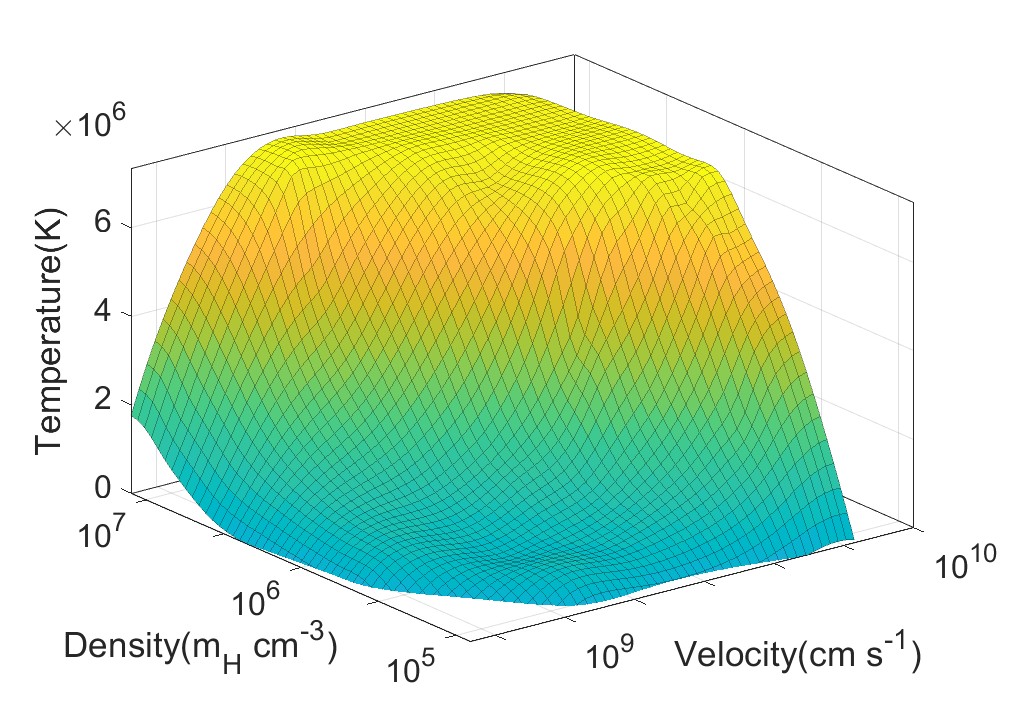}
	\caption{Upper panel: energy transfer efficiency under different outflow densities and velocities. Lower panel: effective temperature under different outflow densities and velocities. Cloud density is set to be $1\times 10^{10}~{\rm m_{H}~cm^{-3}}$.}
    \label{figure_a1}
\end{figure}

To investigate the parameter dependence of efficiency and effective temperature, we vary the outflow velocity from 0.02 $c$ to 0.2 $c$, outflow density from $1\times 10^5~{\rm m_{H}~cm^{-3}}$ to $3\times 10^7~{\rm m_{H}~cm^{-3}}$, cloud density is set to be $1\times 10^{10}~{\rm m_{H}~cm^{-3}}$. The results are plotted in Figure \ref{figure_a1}. The efficiency profile can be described by a polynomial:
\begin{equation}
  \begin{aligned}
    \eta = &a_1+a_2{x}+a_3{y}+a_4{x^2}+a_5{xy}+a_6{y^2}\\
           &+a_7{x^3}+a_8{x^2y}+a_9{xy^2}+a_{10}{y^3}
  \end{aligned}
\end{equation}
where $x=(\log v_{\rm w})/9.5$, $y=(\log\rho_{\rm w})/6.2$, $a_1=2.2\times10^{-2}$, $a_2=-1.7\times10^{-2}$, $a_3=3.0\times10^{-3}$, $a_4=1.2\times10^{-2}$, $a_5=9.1\times10^{-3}$, $a_6=9.8\times10^{-3}$, $a_7=5.5\times10^{-3}$, $a_8=1.4\times10^{-2}$, $a_9=8.0\times10^{-3}$, $a_{10}=3.8\times10^{-3}$.

The effective temperature profile can be described by a polynomial:
\begin{equation}
  \begin{aligned}
        \log{T} = &a_1 + a_2{x} + a_3{y} + a_4{x^2} + a_5{xy} + a_6{y^2} + a_7{x^3}\\
                  &+ a_8{x^2y} + a_9{xy^2} + a_{10}{y^3}
  \end{aligned}
\end{equation}
where $x=(\log v_{\rm w})/9.4$, $y=(\log\rho_{\rm w})/6.2$, $a_1=6.7$, $a_2=0.29$, $a_3=0.28$, $a_4=-5.8\times 10^{-2}$, $a_5=-0.21$, $a_6=-9.6\times 10^{-2}$, $a_7=1.6\times10^{-2}$, $a_8=-2.9\times10^{-2}$, $a_9=3.5\times10^{-2}$, $a_{10}=5.5\times10^{-3}$.

\begin{figure}
	\includegraphics[width=\columnwidth]{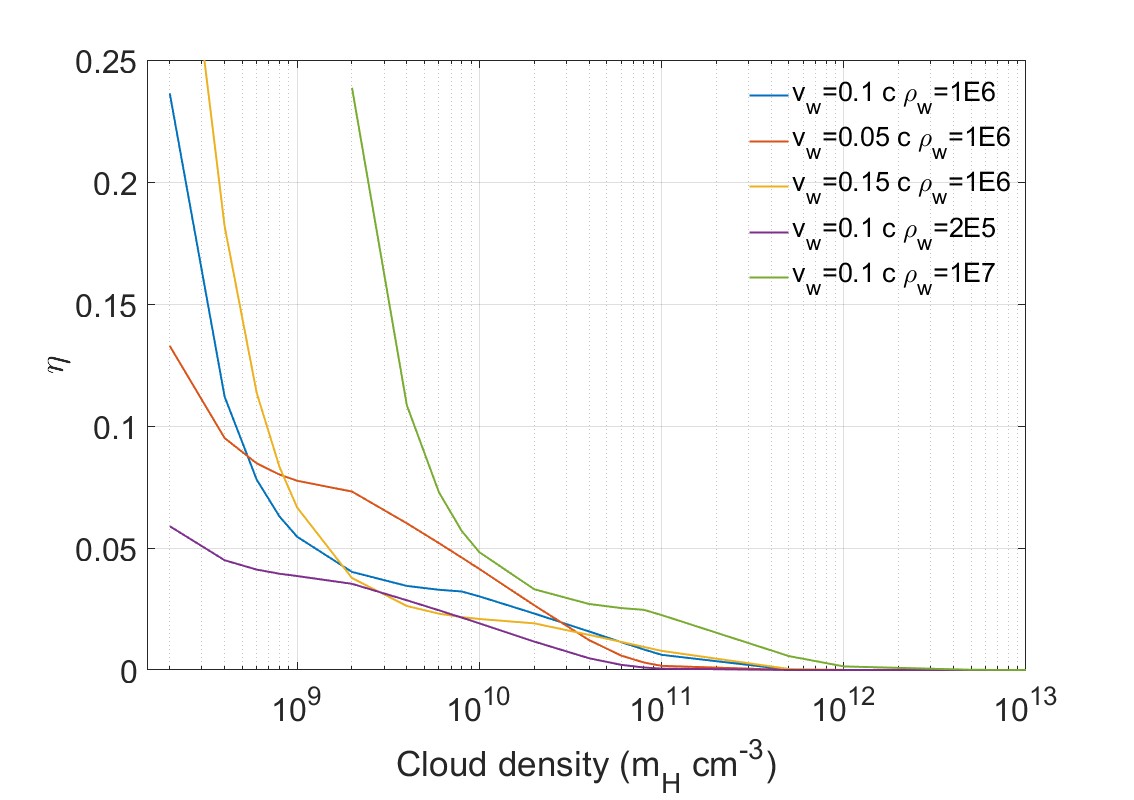}
	\caption{Energy transfer efficiency under different cloud densities.}
    \label{figure_a2}
\end{figure}

Besides, we vary the cloud density from $1\times 10^8~{\rm m_{H}~cm^{-3}}$ to $3\times 10^{13}~{\rm m_{H}~cm^{-3}}$ and keep outflow velocity and density unchanged. The results are plotted in Figure \ref{figure_a2}. The efficiency profile of outflow velocity $v_{\rm w}=0.1~c$ and density $\rho_{\rm w}=1\times 10^6~{\rm m_{H}~cm^{-3}}$ can be described by:
\begin{equation}
    \eta = a_1{\rm exp}(-((x-b_1)/c_1)^2) + a_2{\rm exp}(-((x-b_2)/c_2)^2)
\end{equation}
where $x=\log\rho_{\rm c}$, $a_1=37.4$, $b_1=4.5$, $c_1=1.7$, $a_2=3.2\times10^{-2}$, $b_2=9.6$, $c_2=1.1$.

%%%%%%%%%%%%%%%%%%%%%%%%%%%%%%%%%%%%%%%%%%%%%%%%%%

% Don't change these lines
\bsp	% typesetting comment
\label{lastpage}
\end{document}